\documentclass[twocolumn]{aastex6}
\RequirePackage{lineno}
\usepackage{amsmath}
\usepackage{amssymb}
\usepackage{amsthm}
\usepackage{natbib,aasdefs,url,bm}
\usepackage{array}
\usepackage{float}
\usepackage{graphicx}
\usepackage{subfigure}
\usepackage{color}

\newcounter{ichi}
\newcounter{ni}
\newcounter{san}
\newcounter{yon}
\def\be{\begin{equation}}
\def\ee{\end{equation}}
\def\ba{\begin{eqnarray}}
\def\ea{\end{eqnarray}}

\newcommand{\ergcms}{\mbox{erg~cm$^{-2}$ s$^{-1}$}}
\newcommand{\gray}{${\gamma}$-ray}

\slugcomment{}

\shorttitle{Fast Radio Bursts with Extended Gamma-Ray Emission?}
\shortauthors{Murase, M\'esz\'aros, \& Fox}

\begin{document}

\title{Fast Radio Bursts with Extended Gamma-Ray Emission?}
\author{Kohta Murase\altaffilmark{1,2,3,4}, Peter M\'esz\'aros\altaffilmark{1,2,3}, and Derek B. Fox \altaffilmark{2,3}}
\altaffiltext{1}{Department of Physics, The Pennsylvania State University, University Park, PA 16802, USA}
\altaffiltext{2}{Department of Astronomy \& Astrophysics, The Pennsylvania State University, University Park, PA 16802, USA}
\altaffiltext{3}{Center for Particle and Gravitational Astrophysics, The Pennsylvania State University, University Park, PA 16802, USA}
\altaffiltext{4}{Yukawa Institute for Theoretical Physics, Kyoto University, Kyoto 606-8502, Japan}


\begin{abstract}
We consider some general implications of bright \gray~counterparts to fast radio bursts (FRBs).  We show that even if these manifest in only a fraction of FRBs, \gray~detections with current satellites (including {\it Swift}) can provide stringent constraints on cosmological FRB models. If the energy is drawn from the magnetic energy of a compact object such as a magnetized neutron star, the sources should be nearby and be very rare. If the intergalactic medium is responsible for the observed dispersion measure, the required \gray~energy is comparable to that of the early afterglow or extended emission of short \gray~bursts. While this can be reconciled with the rotation energy of compact objects, as expected in many merger scenarios, the prompt outflow that yields the \gray s is too dense for radio waves to escape. Highly relativistic winds launched in a precursor phase, and forming a wind bubble, may avoid the scattering and absorption limits and could yield FRB emission. Largely independent of source models, we show that detectable radio afterglow emission from \gray~bright FRBs can reasonably be anticipated. Gravitational wave searches can also be expected to provide useful tests. 
\end{abstract}


\section{Introduction}
Fast radio bursts (FRBs) are short radio transients~\citep{Lorimer+07,Keane+12,Thornton+13}, and 
have recently become the object of much scrutiny~\citep{Keane16-frbrev,2016MPLA...3130013K}.
Their inferred rate is ${\cal R}_{\rm FRB} \sim 10^{-3}  \ \rm yr^{-1}~galaxy^{-1}$~\citep{Thornton+13}, or about $10\%$ of supernovae, ${\cal R}_{\rm SN} \sim 10^{-2} \ \rm yr^{-1}~galaxy^{-1}$, which is much higher than that of \gray~bursts (GRBs), ${\cal R}_{\rm GRB} \sim 10^{-6}-10^{-5} \ \rm yr^{-1}~galaxy^{-1}$~\citep{2010MNRAS.406.1944W}.
The dispersion measures (DMs), ${\rm DM} \gtrsim 500 \ \rm cm^{-3} pc$, suggest that they are
cosmological, with redshifts $z \sim 0.5\mbox{-}1$~\citep{NE2001a}.  
The typical flux is $F_{\nu} \sim 0.1-1~\rm Jy$, implying a total emitted energy of $\sim 10^{38}-10^{42}
\ \rm erg$ at cosmological distances, if the emission is isotropic. 
The durations are $\delta t \lesssim 5 \ \rm ms$, indicating relatively compact emission regions, $c \delta t \lesssim 1500 \ \rm km$, if they are non-relativistic. So far all FRBs but one have been non-repeating, so a significant fraction of them may be one-time events.  
The exception, FRB 121102~\citep{2016Natur.531..202S}, could indicate either more than one FRB type, or a difference in environments~\citep{Dai+16frbrep}.

Their origin has been under intense debate. Cosmological FRBs require strong radio emission, and most models involve a neutron star (NS) or a black hole (BH), and the collapse of an accreting NS to a BH has also been discussed, as well as NS-NS/NS-BH/BH-BH binary mergers~\citep[see recent reviews][and references therein]{Keane16-frbrev} (note that the latter merger models could not explain repeating events from the same system).  The most common models involve a young magnetar~\citep[][]{Popov&Postnov10} or giant pulses from a fast-rotating neutron star~\citep[e.g.,][]{2014A&A...562A.137F,Cordes&Wasserman16}.  On the other hand, various merger models, which could be coincident gravitational waves or GRBs, have been invoked~\citep[e.g.,][]{2013PASJ...65L..12T,Kashiyama+13,2014ApJ...780L..21Z,2015ApJ...814L..20M,2016ApJ...822L..14Z}. 
Such systems could be expected to leave an afterglow~\citep[e.g.,][]{2014PASJ...66L...9N,2014ApJ...792L..21Y,MKM16}, and searches have been underway. The first claim of a counterpart was of a long-lasting radio afterglow leading to a host galaxy determination \citep{Keane+16}; however, the host was subsequently found to have re-brightened \citep{Williams&Berger16,Vedantham+16frbag} indicating that it was a flaring galaxy.  More recently, a persistent radio counterpart and the host galaxy of FRB 121102 were discovered~\citep{2017arXiv170101098C,2017ApJ...834L...8M,2017ApJ...834L...7T}, and its emission properties are consistent with the theoretical predictions for pulsar-driven supernova remnants~\citep{MKM16}. However, FRB 121102 is the only repeating FRB, so that the origin of FRBs is still an open question.  

Searches for \gray~counterparts have been going on for quite a while without significant success~\citep{2014ApJ...790...63P,2016ApJ...827...59T,2016MNRAS.460.2875Y}, until the recent announcement~\citep{DeLaunay+16frb131104} of a coincidence between FRB 131104 and a {\it Swift}-BAT ~\gray~transient.  Based on the radio DM, the redshift is $z \sim 0.55$, corresponding to a luminosity distance $d_L\sim3$~Gpc, assuming the $\Lambda$CDM cosmology. This detection of a possible gamma-ray counterpart will give us profound implications, if it is not accidental.  \cite{DeLaunay+16frb131104} determined for it an isotropic-equivalent \gray~energy of ${\mathcal E}_\gamma\approx5\times{10}^{51}$~{\rm erg} with an observed duration of $\approx377$~s.
The confidence is only $3.2 \sigma$, and \gray-bright FRBs should be very rare. But it can be used as the first example where important parameters of possible models can be put to the test. 

The successful detection of an X/\gray~counterpart implies that FRBs may be energetic events like GRBs, if they are cosmological. The sensitivity of the {\it Swift}-BAT imaging trigger is $F_{\rm lim}\sim{10}^{-8}~\ergcms$, which enhances the chance of discovering long-duration X/\gray~transients such as shock breakouts and tidal disruption events, and only a fraction (not all) of the FRBs could be associated with them~\citep{DeLaunay+16frb131104}.  If every GRB does not has an associated FRB, the fraction should be even smaller.

In this work, we consider some general implications that can be obtained from successful detections of an FRB's counterpart.  Our aim is to clarify the required physical conditions and to show the power of a possible \gray~counterpart signal for testing the existing FRB models.  In Sections~2 and 3, we consider \gray~constraints on magnetar-like models and merger-motivated models.  In Section~4 we discuss implications for the afterglow emission.  Throughout this work we use the notation $Q={10}^xQ_x$ in CGS units, unless noted otherwise.

\section{Constraints on Magnetic Burst Scenarios}
\label{sec:magscen}

Among various FRB models, one of the most widely discussed possibilities is the magnetar scenario, where FRBs are attributed to hyper-flares from young magnetars~\citep[e.g.,][]{Popov&Postnov10,Lyubarsky14,MKM16}. If the X/\gray s come from the magnetic energy trapped in a NS, its isotropically equivalent energy is limited by
\be
{\mathcal E}_{\gamma}\lesssim{\mathcal E}_{\rm mag}f_b^{-1}\approx\frac{1}{6}B_c^2R_*^3\simeq1.7\times10^{49}~{\rm erg}~B_{c,15.5}^2R_{*,6}^3f_{b,-1}^{-1},
\ee
where $f_b=\Omega_b/(4\pi) \leq 1$ is the possible beaming factor (for one-side beam), $B_c$ is the internal magnetic field, and $R_*$ is the stellar size. Note that we have ${\mathcal E}_{\rm mag}\simeq2.1\times10^{45}~{\rm erg}~B_{c,9.5}^2R_{*,8.7}^3$ in white dwarf models~\citep{Kashiyama+13}. 
For a \gray~sensitivity $F_{\rm lim}\sim{10}^{-8}~\ergcms$~\citep{Barthelmy+05,2013arXiv1308.3720L}, a successful detection of \gray~counterparts implies that the luminosity distance must be
\be
d_L{(1+z)}^{-1/2}\lesssim180~{\rm Mpc}~{\mathcal E}_{\rm mag,48}^{1/2}f_{b,-1}^{-1/2}F_{\rm lim,-8}^{-1/2}{\Delta T}_{\gamma,2.4}^{-1/2},
\label{distance}
\ee
where ${\Delta T}_{\gamma}$ is the intrinsic \gray~duration in the cosmic rest frame.  Thus, if FRBs are powered by compact objects with strong magnetic fields, they cannot be at gigaparsec distances, unless the \gray s are beamed.  

The DM may be used to infer the distance to FRBs. For example, an FRB with ${\rm DM}=779~{\rm pc}~{\rm cm}^{-3}$ indicates that $d_L\sim3$~Gpc. However, the DM may be dominated by the local environments, including the immediate environments (such as pulsar wind nebulae and supernova remnant ejecta) as well as the ionized plasma in host galaxies~\citep[e.g.,][]{Masui+15,Kulkarni+14,Connor+16,2016ApJ...824L..32P}.  In particular, a non-negligible contribution could come from the supernova ejecta~\citep[but see also][]{Katz16,MKM16}. Although the unshocked ejecta of very young supernova remnants are mostly neutral (depending on their evolution and composition), assuming a mean atomic number $\bar{A}\sim10$ and effective charge $\bar{Z}\sim \bar{A}/2$ for the singly ionized state, the free electron density is estimated to be $n_e=3M_{\rm ej}/(4\pi R_{\rm ej}^3\mu_em_H)$ with $\mu_e^{-1}\sim\bar{A}^{-1}\sim0.1$, and the DM is evaluated as ${\rm DM}\approx n_eR_{\rm ej}\approx n_eV_{\rm ej}T_{\rm age}$. 

On the other hand, free-free absorption in an ionized plasma prevents radio emission from escaping the system. The free-free optical depth is given by
\be
\tau_{\rm ff}\approx8.4\times{10}^{-28}~{\mathcal T}_{e,4}^{-1.35}\nu_{10}^{-2.1}\int dr~n_en_i\bar{Z}^2,
\ee
where ${\mathcal T}_e$ is the electron temperature. Imposing $\tau_{\rm ff}\lesssim1$ gives an upper limit on the DM due to the immediate environment.  Assuming a supernova ejecta with mass $M_{\rm ej}=10~M_\odot~M_{\rm ej,1}$, we obtain
\be
{\rm DM}\lesssim590~{\rm pc~{\rm cm}^{-3}}~\frac{{\mathcal T}_{e,5}^{27/50}\nu_{9}^{21/25}M_{\rm ej,1}^{1/5}{(\bar{A}/10)}^{2/5}}{{(\bar{Z}/5)}^{4/5}\mu_{e,1}^{3/5}},
\ee
which is consistent with previous results, ${\rm DM}\lesssim21~{\rm pc}~{\rm cm}^{-3}~{\mathcal T}_{e,2.5}^{27/50}\nu_{9}^{21/25}~M_{\rm ej,0.5}^{1/5}{(\bar{A}/10)}^{2/5}{(\bar{Z}/5)}^{-4/5}\mu_{e,1}^{-3/5}$ \citep{MKM16}. 
Correspondingly, for the supernova ejecta to give a significant contribution ${\rm DM}\gtrsim300~{\rm pc}~{\rm cm}^{-3}$, the age should satisfy
\ba
40~&{\rm yr}&~{\mathcal T}_{e,5}^{-27/100}\nu_{9}^{-21/50}M_{\rm ej,1}^{2/5}{(\bar{Z}/5)}^{2/5}{(\bar{A}/10)}^{-1/5}\mu_{e,1}^{-1/5}\nonumber\\
& &V_{\rm ej,8.5}^{-1} \lesssim T_{\rm age} \lesssim55~{\rm yr}~M_{\rm ej,1}^{1/2}\mu_{e,1}^{-1/2}V_{\rm ej,8.5}^{-1}.
\label{freefree}
\ea
In principle, it is possible to expect that the DM is dominated by the supernova ejecta.  However, the allowed parameter space is rather limited, so it should be a rare event.  This constraint is applied independently of \gray~observations, if the supernova ejecta significantly contribute to the DM.  

Although there is large uncertainty, the apparent cosmological FRB rate has been estimated to be $\rho_{\rm FRB}\sim{10}^4~{\rm Gpc}^{-3}~{\rm yr}^{-1}$ with large uncertainty~\citep[e.g.,][]{Kulkarni+14,2016MNRAS.460L..30C,2016ApJ...825L..12C}. Equation~(\ref{distance}) suggests that \gray s are detectable for $d\lesssim100-200$~Mpc in the magnetar scenario. Thus, even if all FRBs are intrinsically accompanied by \gray s, only a fraction of FRBs have \gray~counterparts that can be detected by {\it Swift}-BAT.  Indeed, the all-sky rate within $\sim100~{\rm Mpc}^{-3}$ is $\sim40~{\rm yr}^{-1}$. The \gray~detectability depends on detectors' sensitivities and fields of view. If all FRBs were like FRB 131104, one would obtain $\sim25~{\rm yr}^{-1}$ above the BAT threshold~\citep{DeLaunay+16frb131104}, which is at most comparable to the above number within large uncertainties.  In addition, Equation~(\ref{freefree}) suggests that most magnetars will not have required DM values, so that FRBs accompanied by \gray s should be even more rare. 

In summary, because of the energetics argument, the magnetar scenario predicts that \gray~counterparts of FRBs can be detected only for nearby events, so that the DM should be largely attributed to local environments. The allowed parameter space is so narrow that FRBs with detectable \gray~counterparts and large DMs are difficult in the magnetar scenario.

\section{Gamma-Ray Constraints on Relativistic Outflow Scenarios}
\label{sec:radiocomp}

\subsection{Implications for Compact Merger Scenarios}

GRBs are known energetic transients, but the observed GRB rate is much lower than the FRB rate.  Observationally, only a fraction of the FRBs may still be accompanied by \gray s, but this could naturally occur if \gray s are strongly beamed, whereas radio emission is more isotropic. 
On the other hand, compared to ordinary burst emission, the association with longer-duration transients is less constrained. Interestingly, some short GRBs (SGRBs) have showed extended emission in the X-ray and \gray~ranges~\citep[e.g.,][and references therein]{2013MNRAS.431.1745G,2015MNRAS.452..824K}, which seems to be consistent with {\it Swift}-BAT's observation of FRB 131104~\citep{DeLaunay+16frb131104}.

Two popular possibilities have been considered as energy sources for the SGRB extended emission. The first is the rotation energy of a rapidly rotating magnetar~\citep{Blackman_Yi_1998,1998A&A...333L..87D,Zhang_Meszaros_2001}. Numerical relativity simulations of NS-NS mergers have suggested that a hypermassive magnetar may be formed as a remnant of the merger~\citep{2005PhRvL..94t1101S}.  Its lifetime depends on the equation of state, which is currently uncertain. If the equation of state is not very stiff, the proto-NS collapses into a BH after the NS loses its angular momentum.  However, in principle, such a magnetar could survive for a long time, and it has been speculated that this explains the extended emission~\citep{2006Sci...311.1127D,2006MNRAS.372L..19F,2008MNRAS.385.1455M}.  
Using the results of magnetohydrodynamic simulations~\citep{Gruzinov_2005,Spitkovsky_2006}, the spin-down energy-loss rate is estimated to be
\ba
L_{\rm sd}&\approx&\frac{20\pi^4B_*^2R_*^6{P_i}^{-4}}{3c^3}\nonumber\\
&\simeq&2.4\times10^{49}~{\rm erg}~{\rm s}^{-1}~B_{*,15}^2P_{i,-3}^{-4}R_{*,6}^6
\ea
and the spin-down time is
\be
T_{\rm sd}\approx\frac{3P_i^2{\mathcal I}c^3}{10\pi^2B_*^2R_*^6}\simeq800~{\rm s}~B_{*,15}^{-2}P_{i,-3}^{2}R_{*,6}^{-4},
\ee
where $P_i$ is the initial rotation period and $B_{*}$ is the dipole magnetic field at the surface. 
To have an extended emission with duration ${\Delta T}_\gamma$, the spin-down time should satisfy $T_{\rm sd}\gtrsim{\Delta T}_\gamma$, where ${\Delta T}_\gamma$ may be attributed to the time of the BH formation~\citep{2014A&A...562A.137F,2014ApJ...780L..21Z}.  With $L_{\rm sd}/f_b=L_w\sim{10}^{49}~{\rm erg}~{\rm s}^{-1}$ and ${\Delta T}_\gamma\sim10^{2.4}$~s, solving these leads to
\ba
B_{*}&\gtrsim&2.8\times{10}^{16}~{\rm G}~L_{w,49}^{-1/2}{\Delta T}_{\gamma,2.4}^{-1}R_{*,6}^{-1}f_{b,-1.5}^{-1/2}\nonumber\\
P_i&\lesssim&{16}~{\rm ms}~L_{w,49}^{-1/2}{\Delta T}_{\gamma,2.4}^{-1/2}R_{*,6}f_{b,-1.5}^{-1/2},
\ea
so rather strong magnetic fields \citep[compared to surface magnetic fields indicated for Galactic magnetars][]{2008A&ARv..15..225M} may be required, although the exact values are affected by uncertainties in, e.g., effects of higher multipoles, the proto-NS size, and the angle between rotation and magnetic axes. 

Another possibility is the fallback accretion and subsequent energy extraction from a spinning BH~\citep[e.g.,][]{2014ApJ...796...13N,2015ApJ...804L..16K}.  The rotation energy of a Kerr BH can be extracted via the Blandford-Znajek process~\citep{1977MNRAS.179..433B}, and its absolute jet power is estimated to be~\citep[e.g.,][]{1999ApJ...523L...7A,2011MNRAS.418L..79T}
\begin{eqnarray}
L_{\rm BZ}&\approx& \frac{1}{8\pi c}\Omega_H^2 B_p^2 R_H^4\nonumber\\
&\simeq&6.9\times{10}^{47}~{\rm erg}~{\rm s}^{-1}~B_{p,13.5}^2M_{\rm BH,0.5}^2,
\end{eqnarray}
where $\Omega_H$ is the BH rotation frequency, $R_H$ is the horizon radius, and $B_p$ is the anchored field.
The rotation energy of the BH is estimated to be ${\mathcal E}_{\rm rot}=[1-{(1/2+\sqrt{1-a^2}/2)}^{1/2}]M_{\rm BH}c^2\sim4\times{10}^{53}~{\rm erg}~M_{\rm BH,0.5}$, where $a\sim0.7$ is the dimensionless Kerr parameter.  As long as ${\mathcal E}_{\rm rot}/L_{\rm BH}$ is large enough, the duration is determined by the fallback accretion time scale that is 
\ba
T_{\rm BZ}\sim300~{\rm s}~B_{p,13.5}^{-6/5}M_{\rm BH.0.5}^{-6/5}T_{\rm vis,-1}^{2/5}M_{\rm fb,-2.5}^{3/5}, 
\ea
where $T_{\rm vis}$ is the viscous time scale and $M_{\rm fb}$ is the fallback mass~\citep{2015ApJ...804L..16K}.  Thus, this scenario is also a viable option.  

Although all models seem speculative, the association with orphan SGRBs is an appealing possibility~\citep{DeLaunay+16frb131104} since compact merger scenarios (including NS-NS/NS-BH/BH-BH models) predict that gravitational waves associated with FRBs can be detected by Advanced LIGO.  The detection of GW signals can prove that some FRBs come from mergers.  By observing the timing among FRB, GW, and associated \gray~emissions, one could probe the emission region of FRBs as well as the emission mechanism of extended \gray~emission, which is especially relevant to test the precursor model described in the next section. 

However, there are two general remarks. First, one should note that the rate of FRBs with such large radiation energy must be much smaller than that of FRBs. Second, to be consistent with {\it Swift}-BAT observations, the rate of orphan SGRBs showing extended emission cannot be much larger than the SGRB rate, and the former beaming factor cannot be not much larger than that of SGRBs, $f_{b}\sim0.01$~\citep{DeLaunay+16frb131104}.  Note that the observed SGRB rate is $\rho_{\rm SGRB}\sim{10}~{\rm Gpc}^{-3}~{\rm yr}^{-1}$~\citep[e.g.,][]{2012MNRAS.425.2668C,2015ApJ...815..102F}, and it has been thought that $\sim20-30$\% of SGRBs may have the extended emission~\citep{2015MNRAS.452..824K}. 
Obviously, the apparent rate of SGRBs is much smaller than the total rate of FRBs, $\rho_{\rm FRB}\sim10^4~{\rm Gpc}^{-3}~{\rm yr}^{-1}$, so that only a fraction of total FRBs could be accompanied by \gray s. For example, \gray~emission may be beamed, while FRB emission may be more isotropic.

\subsection{Radio Compactness Problem}

Let us assume a generic relativistic outflow with isotropic-equivalent outflow luminosity, $L_w$, and bulk Lorentz factor, $\Gamma_0$.  If \gray s are detected by current satellites such as {\it Swift}, it implies $L_w\gg L_{\rm FRB}$, and the outflow would not be so tenuous.  As in X-ray flares, long-duration \gray~emission is likely to have an internal dissipation origin, so we assume the late prompt emission model~\citep[e.g.,][]{2007ApJ...658L..75G,2011ApJ...732...77M}, in which the extended X/\gray~emission is produced by internal shocks or magnetic reconnections via synchrotron radiation~\citep[see also][for the case of the photospheric scenario]{2014ApJ...796...13N}.  In the internal shock model, internal dissipation occurs at $r\sim\Gamma_0^2c{\Delta t}_\gamma$, where ${\Delta t}_\gamma\leq {\Delta T}_\gamma$ is the variability time. 
The spectrum of the observed extended emission does not look purely thermal. Although we do not exclude the dissipative photosphere model as an option~\citep{2005ApJ...628..847R}, internal dissipation should occur beyond/around the photosphere, whose radius is
\be
r_{\rm ph}\simeq3.7\times{10}^{11}~{\rm cm}~\zeta_eL_{w,49}\Gamma_{0,1.5}^{-3}{[(1+\sigma)]}^{-1}.
\ee 
where $n'_{w}\approx L_w/[4\pi r^2 \Gamma_0^2 m_pc^3(1+\sigma)]$ is the comoving baryon density (here $\sigma$ is the magnetization parameter) and $\zeta_e(\leq m_p/m_e)$ is a possible enhancement factor due to electron-positron pairs. Note that the acceleration of the outflow is assumed to cease at sub-photospheres, and we may use $\tau_T\approx\zeta_e n'_{w}\sigma_T(r/\Gamma_0)$. 

On the other hand, the dissipation radius at which internal dissipation (including FRB emission) occurs, is limited as $r<{\rm max}[r_{\rm dec},r_{\rm BM}]$, where
\ba
r_{\rm dec}&\approx&{[3{\mathcal E}/(4\pi nm_pc^2\Gamma_0^2)]}^{1/3}\nonumber\\
&\simeq&2.0\times{10}^{17}~{\rm cm}~{\mathcal E}_{51.7}^{1/3}\Gamma_{0,1.5}^{-2/3}n_{-1}^{-1/3}
\ea
is the deceleration radius, and 
\be
r_{\rm BM}\simeq1.4\times{10}^{17}~{\rm cm}~{\mathcal E}_{51.7}^{1/4}n_{-1}^{-1/4}{\Delta T}_{\gamma,2.4}^{1/4}, 
\ee
is given by the Blandford-McKee self-similar solution. 

In principle, FRB emission could occur during the extended \gray~emission. However, one sees that this is challenging due to the radio compactness problem.  First of all, the emission region should be outside the photosphere, otherwise the radio emission is diminished by the Thomson scattering.  

Second, the synchrotron absorption due to non-thermal electrons producing the extended emission is relevant~\citep{MKM16,2016ApJ...819L..12Y}, unless an inverse-population of electrons is formed.  The absorption frequency, at which $\tau_{\rm sa}(\nu_{\rm sa})=1$, is estimated by equating the intrinsic surface flux to a blackbody surface flux as
\be
2 k{{\mathcal T}_e}'\frac{{\nu'}_{\rm sa}^2}{c^2}\approx2 \frac{{L_{\nu_{\rm sa}}}}{4\pi r^2\Gamma_0}.
\ee 
For simplicity, let us assume a simple power-law spectrum, $L_E/E \propto E^{-\alpha}$ at $E=h\nu<E^b$ and $k{{\mathcal T}_e}'=\gamma'_{ea}m_ec^2$, where $E^b$ is the break energy in the burst frame. For $\alpha=1$, which is typical in GRB prompt emission and consistent with the X/\gray~observation of FRB 131104~\citep{DeLaunay+16frb131104}, we obtain $\nu_{\rm sa}\simeq1.0\times{10}^{12}~{\rm Hz}~L_{w,49}^{1/2}\Gamma_{0,1.5}^{2/5}\epsilon_B^{1/10}{(E^b/100~{\rm keV})}^{-2/5}r_{16}^{-1}$. Then, requiring $\tau_{\rm sa}(\nu)<1$ gives
\be
r\gtrsim1.0\times{10}^{19}~{\rm cm}~\nu_9^{-1}L_{w,49}^{1/2}\Gamma_{0,1.5}^{2/5}\epsilon_B^{1/10}{(E^b/100~{\rm keV})}^{-2/5}.
\ee
Thus, radio emission cannot escape from a late prompt outflow producing the synchrotron \gray~emission. The \gray~observation suggest a large value of $L_w$, which leads to a strong constraint given that the \gray s originate from relativistic electrons. Note that in general, details depend on the photon spectrum or the electron distribution, but our assumption is satisfied in the sufficiently fast cooling case. 
  
Another constraint can be placed by the induced-Compton scattering~\citep[e.g.,][]{1970SvPhU..13..307Z}.  First, we consider a region with a comoving size of $r/\Gamma_0$. Assuming an isotropic emission in this frame, its optical depth is~\citep{1971Ap&SS..13...56M,2008ApJ...682.1443L}
\be
\tau_{\rm ic}\approx\frac{3\zeta_e n'_{w}\sigma_TI'_{\rm FRB}r}{2\nu'^2\gamma'_{T}m_e\Gamma_0},
\label{ic1}
\ee
where $I'_{\rm FRB}$ is the radio intensity and $\gamma'_Tm_ec^2$ is the thermal energy of electrons.
Imposing $\tau_{\rm ic}\lesssim10$ gives
\begin{eqnarray}
r&\gtrsim&6.3\times{10}^{17}~{\rm cm}~\frac{L_{\rm FRB, 43}^{1/3}{L}_{w,49}^{1/3}\zeta_e^{1/3}}{{\Gamma}_{0,1.5}^{2/3}\nu_9{\gamma'_{T}}^{1/3}{(1+\sigma)}^{1/3}},
\end{eqnarray}
which is typically larger than $r_{\rm dec}$. 

The above constraint is strong but rather conservative. The FRB emission has a short pulse of $\delta t\sim1~{\rm ms}$, so we may consider a compact blob with $l'_{c}\approx\Gamma_0c\delta t(\lesssim r/\Gamma_0)$ (in the outflow comoving frame).  
Then, the induced-Compton scattering optical depth is estimated to be~\citep{1978MNRAS.185..297W,2007ApJ...658L...1M,2013PTEP.2013l3E01T}
\be
\tau_{\rm ic}\approx \zeta_e n'_{w}\sigma_Tl'_c\left(\frac{3k{\mathcal T'}_{b}}{\gamma'_{T}m_ec^2}\right),
\label{ic2}
\ee
where ${\mathcal T'}_b$ is the brightness temperature given by ${\mathcal T}'_b=c^2 I'_{\rm FRB}/(2k\nu'^2)$. 
For a spherical blob, using the isotropic-equivalent FRB luminosity ($L_{\rm FRB}$), we obtain $I'_{\rm FRB}\approx L_{\rm FRB}/(8\pi^2 \Gamma_0^3 {l'_c}^2\nu)$ and ${\mathcal T}'_b\simeq1.5\times{10}^{31}~{\rm K}~L_{\rm FRB,43}\Gamma_{0,1.5}^{-3}\nu_9^{-3}{\delta t}_{-3}^{-2}$.  
If we assume that the emission occurs at $r\approx\Gamma_0^2c\delta t$, the Lorentz factor is constrained to be
\be
\Gamma_0>\Gamma_c\simeq1.8\times{10}^4~\frac{L_{\rm FRB,43}^{1/8}L_{w,49}^{1/8}\zeta_e^{1/8}}{\nu_9^{3/8}{\delta t}_{-3}^{3/8}{\gamma'_{T}}^{1/8}{(1+\sigma)}^{1/8}}.
\ee
Thus, the radio emission could avoid induced-Compton scatterings if $\Gamma_0$ exceeds the critical Lorentz factor ($\Gamma_c$), at which Equations~(\ref{ic1}) and (\ref{ic2}) are equal.  Or, the constraints can be weaker if the flow is Poynting dominated ($\sigma\gg1$) or the plasma is hot ($\gamma'_{T}\gg1$).  


\section{Precursor Burst in a Wind Bubble?}
\label{sec:precbub}

FRB emission during the precursor phase of compact mergers has been discussed by several authors~\citep{2013ApJ...768...63L,2013PASJ...65L..12T,2016ApJ...822L...7W}.  
Although the magnetic fields of binary pulsars are uncertain, it is often thought that one of the two NSs has a weak magnetic field as a recycled pulsar, whereas the other has a strong magnetic field with $B_*\sim{10}^{12}-{10}^{13}$~G.  The merger time due to gravitational wave losses is 
\be
t_m=\frac{5}{512}\frac{c^5a^4}{G^3M^3}\simeq3.0~{\rm ms}~{\left(\frac{a}{10~{\rm km}}\right)}^4, 
\ee
where $a$ is the separation and their mass is assumed to be $1.4~M_\odot$.  
The pre-merger luminosity is uncertain, but the maximum luminosity is estimated to be $L_{\rm pre}^{\rm max}\sim4\times{10}^{45}~{\rm erg}~{\rm s}^{-1}~B_{*,13}^2t_{m,-3}^{-13/8}$~\citep{2012ApJ...757L...3L}.  The isotropic-equivalent wind luminosity could be larger if the flow is collimated as $L_w\approx L_{\rm pre}/f_b$, and $L_w\gtrsim L_{\rm FRB}\sim{10}^{43}~{\rm erg}~{\rm s}^{-1}$ is expected to explain FRBs. The wind magnetic field at radius $r$ may be as strong as $B\approx82~{\rm G}~L_{w,45}^{1/2}r_{15.5}^{-1}$ in the burst frame.  
  
Even in this precursor phase, FRB emission cannot be produced close to the binary.  The light cylinder of the system may be $R_{\rm lc}=cP/(2\pi)\simeq4.8\times{10}^{6}~P_{-3}~{\rm cm}$, which can be comparable to the photospheric radius of the precursor fireball, $r_{\rm ph}\sim3.8\times{10}^6~{\rm cm}~{(r_0/a)}^{1/2}B_{*,13}^{1/2}{(a/30~{\rm km})}^{-9/8}$~\citep{2016MNRAS.461.4435M}, where $r_0$ is the initial radius.  A radio pulse produced in such a dense region would be significantly diminished. 
Note that the similar constraint would be applied to NS-BH binary models~\citep{2016PhRvD..94b3001D}.  

The situation is much relaxed at large radii, given that the wind is magnetically accelerated to a highly relativistic speed. Using Equation~(\ref{ic1}), the induced-Compton scattering limit becomes~\footnote{Note that a geometrical factor should be taken into account for a radio beam that has successfully escaped from the emission region and propagates in the plasma~\citep[e.g.,][and references therein]{1978MNRAS.185..297W,2013PTEP.2013l3E01T}.}
\be
r\gtrsim3.6\times{10}^{14}~{\rm cm}~\frac{L_{\rm FRB, 43}^{1/3}L_{w,45}^{1/3}}{{\Gamma}_{w,6}^{2/3}\nu_9{\gamma'_{T}}^{2/3}{(1+\sigma)}^{1/3}},
\label{preic1}
\ee
where we have assumed $\zeta_e=m_p/(\gamma'_Tm_e)$ and $r\lesssim\Gamma_w^2c \delta t\simeq3.0\times{10}^{19}~{\rm cm}~\Gamma_{w,6}^2{\delta t}_{-3}$. 

At present, there is no convincing model for the coherent mechanism of FRB emission, in particular in cosmological scenarios.  As an alternative possibility, we consider the burst-in-bubble model~\citep{MKM16}.  A highly magnetized impulsive wind, which may be caused by some inhomogeneous energy injection, would run into a cleaner bubble environment created by a previous wind from the progenitor. Then, this relativistic wind pulse leads to significant dissipation as it interacts with the existing nebula.  The initial nebula is expected to be small and dim.  For an old pulsar with $B_{*}\sim{10}^{12.5}$~G and $P\sim10$~s (corresponding to $T_{\rm age}\sim0.3$~Gyr), the spin-down luminosity is estimated to be $L_{\rm sd}\approx2.4\times{10}^{28}~{\rm erg}~{\rm s}^{-1}~B_{*,12.5}^2P_{1}^{-4}R_{*,6}^6$.  Note that old pulsars are in the interstellar medium rather than inside the supernova ejecta, forming a bow shock.  Using a pulsar velocity, $V_k\sim100-1000~{\rm km}~{\rm s}^{-1}$, the typical nebular radius is estimated to be $R_{w}\approx{[L_{\rm sd}/(6\pi nm_pV_{k}^2c)]}^{1/2}\simeq5.0\times{10}^{13}~{\rm cm}~B_{*,12.5}P_{1}^{-2}R_{*,6}^3n_{-1}^{-1/2}V_{k,7}^{-1}$, using the isobaric condition for the nebular pressure, $p_{\rm ts}=(2/3)\Gamma_w^2 n_w mc^2\approx L_{\rm sd}/(6\pi R_{w}^2 c)$, and the ram pressure of the cold interstellar medium, $p_{\rm bs}=nm_pV_k^2$. The nebula is freshly energized only for $\sim R_{\rm ts}/V_k\simeq5.0\times{10}^6~{\rm s}~B_{*,12.5}P_{1}^{-2}R_{*,6}^3n_{-1}^{-1/2}V_{k,7}^{-2}$ and the magnetic field is estimated to be $B_{\rm nb}\simeq75~{\mu~\rm G}~\epsilon_{B,-3}^{1/2}n_{-1}^{1/2}V_{k,7}^{1/2}$. Note that this nebular size may not satisfy Equation~(\ref{preic1}).  

However, the nebula is so tenuous that it would be significantly affected by a continuous energy injection from a precursor outflow.  Although its detailed evolution needs dynamical calculations, the causality and pressure balance conditions imply that the nebula may 
expand up to $\sim2\times{10}^{15}~{\rm cm}~B_{*,12.5}^{16/29}n_{-1}^{-8/29}V_{k,7}^{-16/29}$. If the wind is accelerated to $\Gamma_w\gg{10}^4$, the pressure balance at the contact discontinuity leads to a Lorentz factor of the merged wind of $\Gamma_m\gtrsim1400~L_{w,45}^{1/4}n_{-1}^{-1/4}V_{k,7}^{-1/2}R_{w,15.5}^{-1/2}$.   

The impulsive wind is quickly decelerated after interaction with the bubble, and its energy can be efficiently converted into radiation. Particles in the wind can be boosted by $\sim \Gamma_w/(2\Gamma_m)$, while those in the pre-shocked nebula can be boosted by $\sim\Gamma_m$, respectively. 
An inverted population of particles may form in such a situation.  Although details are uncertain, if the synchrotron absorption cross-section is negative in the relevant energy range~\footnote{\cite{Ghi16} obtained $\gamma_e\lesssim110~{\nu}_9^{1/2}B^{-1/2}$.}, the synchrotron maser mechanism~\citep{1985SSRv...41..215W,1992ApJ...391...73G,2002ApJ...574..861S,Lyubarsky14} may operate and could account for the FRB emission.  For $\Gamma_m<\Gamma_c\simeq1.4\times{10}^4~L_{\rm FRB,43}^{1/8}L_{w,45}^{1/8}\nu_9^{-3/8}{\delta t}_{-3}^{-3/8}{\gamma'_{T}}^{-1/4}{(1+\sigma)}^{-1/8}$, the induced-Compton scattering limit gives 
\be
r\gtrsim3.9\times{10}^{13}~{\rm cm}~\frac{L_{\rm FRB, 43}^{1/2}{L}_{w,45}^{1/2}}{{\Gamma}_{m,3.5}^{2}{\delta t}_{-3}^{1/2}\nu_9^{3/2}{\gamma'}_{T,3.5}{(1+\sigma)}^{1/2}},
\ee
which can be satisfied in this burst-in-bubble scenario. 

If the nebula is hot by virtue of a relativistic particle content, shocked particles energized by the relativistic boost can rapidly lose their energies via synchrotron emission in the high-energy \gray~range at $\sim0.5~\Gamma_{m,3.5}$~TeV. However, such a \gray~flash is detectable only for nearby events~\citep{MKM16}.

\section{Radio Afterglow emission}
\label{sec:radioag}

FRBs associated with \gray s should be caused by energetic events, and this may lead to external shocks driven by a high-velocity outflow.  
First, let us consider a relativistic, late prompt outflow, as considered in merger scenarios~\citep[e.g.,][]{2016ApJ...829..112L}.  In the thin shell case, the crossing time is comparable to $t_{\rm dec}\approx r_{\rm dec}/(4\Gamma_0^2c)\simeq1700~{\rm s}~{\mathcal E}_{51.7}^{1/3}\Gamma_{0,1.5}^{-8/3}n_{-1}^{-1/3}$, where ${\mathcal E}$ is the isotropic-equivalent kinetic energy and $n$ is the ambient density. The Lorentz factor at $t>t_{\rm dec}$ is
\begin{equation}
\Gamma\approx{\left(\frac{17\mathcal E}{1024\pi nm_pc^5t^3}\right)}^{1/8}\simeq7.1~{\mathcal E}_{51.7}^{1/8}n_{-1}^{-1/8}t_{5}^{-3/8}.
\end{equation} 
The post-shock magnetic field is estimated to be $B'={(32\pi\epsilon_B\Gamma^2nm_pc^2)}^{1/2}\simeq8.7\times{10}^{-2}~{\rm G}~{\mathcal E}_{51.7}^{1/8}n_{-1}^{3/8}\epsilon_{B,-2}^{1/2}t_{5}^{-3/8}$, where $\epsilon_B$ is the magnetic energy fraction.  Note that the redshift dependence, which can be taken into account easily, is not explicitly shown here. 

Following the standard external forward shock model~\citep{1993ApJ...405..278M,1997ApJ...476..232M}, the injection Lorentz factor of accelerated electrons is estimated to be $\gamma'_{ei}\approx g_s(\epsilon_e/f_e)(m_p/m_e)\Gamma\simeq370~(\epsilon_{e,-1}/f_{e})(g_s/g_{2.4}){\mathcal E}_{51.7}^{1/8}n_{-1}^{-1/8}t_{5}^{-3/8}$, where $g_s=(s-2)/(s-1)$ for $s>2$, $\epsilon_e$ is the energy fraction of non-thermal electrons, and $f_e(\gtrsim m_e/m_p\sim5\times{10}^{-4})$ is their injection fraction. The accelerated electrons cool on a time scale $t'_{\rm rad}\approx 6\pi m_ec/(\sigma_T {B'}^2\gamma'_e[1+Y])$, where $Y$ is the Compton Y parameter. By the condition $t'_{\rm rad}=r/(\Gamma c)$, we have the cooling Lorentz factor, $\gamma'_{ec}\approx 6\pi m_ec/(\sigma_T{B'}^2 t'[1+Y])\simeq3.6\times{10}^4~{\mathcal E}_{51.7}^{-3/8}n_{-1}^{-5/8}\epsilon_{B,-2}^{-1}{(1+Y)}^{-1}t_{5}^{1/8}$. The acceleration time is given by $t'_{\rm acc}\approx\gamma'_em_ec^2/(eB'c)$.  The condition $t'_{\rm acc}=t'_{\rm rad}$ gives the maximum Lorentz factor of electrons, $\gamma'_{eM}\approx{6\pi e/(\sigma_TB'[1+Y])}^{1/2}\simeq3.9\times{10}^8~{\mathcal E}_{51.7}^{-1/16}n_{-1}^{-3/16}\epsilon_{B,-2}^{-1/4}{(1+Y)}^{-1/2}t_{5}^{3/16}$.

Then, we estimate the injection synchrotron frequency to be $\nu_{i}\approx\Gamma{\gamma'_{ei}}^23eB'/(4\pi m_ec)\simeq3.6\times{10}^{11}~{\rm Hz}~{\mathcal E}_{51.7}^{1/2}t_{5}^{-3/2}\epsilon_{B,-2}^{1/2}\epsilon_{e,-1}^{2}g_{2.4}^{2}f_e^{-2}$, the cooling synchrotron frequency is $\nu_{c}\approx\Gamma{\gamma'_{ec}}^23eB'/(4\pi m_ec)\simeq3.3\times{10}^{15}~{\rm Hz}~{\mathcal E}_{51.7}^{-1/2}n_{-1}^{-1}t_{5}^{-1/2}\epsilon_{B,-2}^{-3/2}{(1+Y)}^{-2}$, and the maximum synchrotron frequency is $\nu_{M}\simeq4.1\times{10}^{23}~{\rm Hz}~{\mathcal E}_{51.7}^{1/8}n_{-1}^{-1/8}t_{5}^{-3/8}{(1+Y)}^{-1}$.  The maximum synchrotron energy flux per frequency is $L_\nu^{\rm max}\approx\Gamma(0.6f_e nr^3)[4\sqrt{3}\pi e^3B'/(3m_ec^2)]\simeq8.2\times{10}^{30}~{\rm erg}~{\rm s}^{-1}~{\rm Hz}^{-1}~{\mathcal E}_{51.7}f_e\epsilon_{B,-2}^{1/2}n_{-1}^{1/2}$.   
A slow-cooling synchrotron spectrum is typically expected, in which the synchrotron spectrum at $\nu<\nu_i$ is $F_{\nu}\propto \nu^{1/3}$, while we expect $F_{\nu}\propto \nu^{1/2-s/2}$ at $\nu_i<\nu<\nu_c$, and $F_{\nu}\propto \nu^{-s/2}$ at $\nu_c<\nu<\nu_M$, respectively.  Here $s$ is the injection spectral index of the accelerated electrons. The radio band typically lies in the range $\nu<\nu_i$, where the synchrotron flux at time $t$ is approximately given by
\be
F_\nu\sim0.09~{\rm mJy}~\nu_9^{1/3}{\mathcal E}_{51.7}^{5/6}n_{-1}^{1/2}\epsilon_{B,-2}^{1/3}\epsilon_{e,-1}^{-2/3}f_e^{5/3}g_{2.4}^{-2/3}t_{5}^{1/2}d_{28}^{-2},\,\,\,\,\,\,\,\,\,\,\,\,
\label{ag1}
\ee
where $s\sim2.4$ is used as a nominal value. Thus, as discussed for GRBs, the radio afterglows are detectable with radio facilities such as the Very Large Array (with a sensitivity of $\sim0.03-0.1$~mJy). 
Note that the self-absorption frequency at $\nu_{\rm sa}\simeq1.2\times{10}^9~{\rm Hz}~{\mathcal E}_{51.7}^{1/5}n_{-1}^{3/5}\epsilon_{B,-2}^{1/5}\epsilon_{e,-1}^{-1}g_{2.4}^{-1}f_e (<\nu_i)$ can be relevant. 

An afterglow emission is promising even in the magnetic burst scenario. Although the ejecta is only mildly relativistic or non-relativistic, there are two advantages. First, in the case of a  \gray~association, the source has to be nearby. Second, the magnetar making a FRB is likely to be so young that the external (supernova ejecta) density is large.  Assuming $d\sim100$~Mpc and ${\mathcal E}\sim{10}^{48}~{\rm erg}$, the non-relativistic afterglow flux at $\nu_m<\nu<\nu_c$ is estimated to be~\citep{MKM16}
\begin{eqnarray}
F_\nu&\sim&5.1~{\rm mJy}~M_{\rm ej,1}^{19/20-s/4}{(V_{\rm ej,8.5}T_{\rm age,9})}^{-57/20+3s/4}\nu_9^{1/2-s/2}\nonumber\\
&\times&{\mathcal E}_{48}^{3/10+s/2}\epsilon_{B,-2}^{1/4+s/4}f_{e,-1.5}{\zeta_e}^{-1+s}t_{6}^{21/10-3s/2}d_{26.5}^{-2}.
\label{ag2}
\end{eqnarray}
Thus, both Equations~(\ref{ag1}) and (\ref{ag2}) imply that the FRB's afterglow signatures are detectable with current radio telescopes carrying out dedicated follow-up observations, but for some FRBs that are accompanied by \gray s. 

The above arguments are generally applicable to various scenarios involving energetic blast waves. As a possible application, let us discuss FRB 131104. Recently, \cite{sr16} reported an upper limit on the radio afterglow flux in the field of view of FRB 131104.  Using $F_\nu<F_{\nu}^{\rm lim}\approx0.07~{\rm mJy}$ at $\nu=5.5$~GHz at $t\sim10^6$~s, we obtain a constraint
\ba
n\epsilon_{B,-2}^{2/3}f_e^{10/3}&\lesssim&2.0\times{10}^{-3}~{\rm cm}^{-3}~\nu_{9.7}^{-2/3}{\mathcal E}_{51.7}^{-5/3}\epsilon_{e,-1}^{4/3}\nonumber\\
&\times&g_{2.4}^{4/3}t_{6}^{-1}{(F_\nu^{\rm lim}/0.07~{\rm mJy})}^{2}.
\ea 
This upper limit is consistent with ambient densities deduced from SGRB afterglows, which are typically $n\sim{10}^{-5}-1~{\rm cm}^{-3}$~\citep[see, e.g., Fig.~9 of][]{2015ApJ...815..102F} (and it also depends on other parameters such as $\epsilon_B$ and $f_e$, which can be low).  While a large parameter space for the standard afterglow scenario for SGRBs is not yet constrained, it disfavors the long GRB afterglow scenario, which is characterized by higher densities, as argued by \cite{sr16}, which is also consistent with the radio compactness issue described above. With the constraints~(\ref{distance}) and (\ref{freefree}), the magnetic burst afterglow scenario is largely excluded. 
These demonstrate how we can use follow-up observations of FRBs with a possible \gray~counterpart.

\section{Summary and Discussion}
\label{sec:disc}

In the previous sections we showed the general importance of \gray~observations {\it for testing FRB models}. Our main points are summarized as follows. 

(i) If the DM is dominated by the intergalactic medium, \gray~detections with {\it Swift}-like detectors imply that FRBs must be energetic events, such as GRBs.  If the energy is supplied by the magnetic energy of a compact object, as in the magnetar hyper-flare model, the source has to be nearby, so the DM should be dominated by local environments.  FRBs with bright \gray~counterparts and large DMs would be difficult in the magnetar scenario, because the allowed age, $T_{\rm age}$, can be largely constrained by the free-free absorption.

(ii) A cosmological interpretation can also be challenged by the {\it radio compactness problem}, which turned out to be usually more severe than the \gray~compactness issue.  
We considered the general implications of late prompt outflows producing extended \gray~emission, which has been observed in some SGRBs~\citep{2015MNRAS.452..824K}, and found that in this case the FRB emission region is unlikely to coincide with the \gray~emission region, especially due to the synchrotron absorption and induced-Compton scattering. 

(iii) On the other hand, FRB emission might occur during the precursor phase of compact mergers. Independently of the \gray~constraints, we showed that a highly relativistic impulsive wind, interacting with a wind bubble, can avoid the induced-Compton scattering constraints. Although we discussed the NS-NS scenario~\citep[e.g.,][]{2013PASJ...65L..12T,2014ApJ...780L..21Z} as an example, the constraints would be applicable to other NS-BH and BH-BH merger scenarios that could work for a given clean environment~\citep{2015ApJ...814L..20M,2016ApJ...822L..14Z}.  

(iv) Based on the standard afterglow model, we showed that radio afterglow emission from \gray~bright FRBs is detectable in merger models, which also encourages future GW searches. We also argued that detectable afterglow emission is expected in the magnetar scenario, following \cite{MKM16}. 

At present, all FRB models that have been proposed in the literature are speculative. Merger scenarios also have caveats.  First, the rate of orphan SGRBs with extended emission cannot be large. The {\it Swift}-BAT observations already imply $f_b\lesssim{10}^{-1.5}$~\citep{DeLaunay+16frb131104}, so the fraction of the FRBs accompanied by \gray s would be at most comparable to the SGRB rate.  Also, merger models would not explain a repeating FRB.  Thus, in this scenario, FRBs should consist of multiple source populations.  On the other hand, even if FRBs may occur in different source populations, there may be a {\it common mechanism} for coherent radio emission, in principle.  Another appealing point is that they have some predictions that can be tested.  If FRBs are produced in the precursor phase, an extended emission should be observed a few seconds later than a radio burst. One also expects that some FRBs show the coincidence of a short \gray~spike a bit after the FRB pulse.  As has been discussed in the literature, even if the internal extended emission is not observed, orphan SGRB afterglows could generically appear as extended, slow-evolving hard X-ray transients (when observed not too far off-axis), as we are just seeing the rise (due to wider beaming) and fall (due to afterglow decay) of the X-ray afterglow.  Radio afterglow observations are also promising, although the detectability and implications depend on ambient densities that can be very low in many merger scenarios.  Another strong test of the compact merger scenarios is the association with GWs, which can be detected by Advanced LIGO, Virgo, and KAGRA, as recently demonstrated by the discovery of BH-BH merger events~\citep{2016PhRvL.116f1102A,2016PhRvL.116x1103A}. 
It has been predicted that NS-NS mergers are detectable up to $\sim200$~Mpc with the designed sensitivity of Advanced LIGO. The field of view of radio telescopes is small, so that coincident detections of FRBs and GWs may be challenging with current facilities.  On the other hand, \gray~detectors have a much larger field of view.  We expect the \gray~coincidence only when we are on-axis, so the detection rate would be at most comparable to that of GWs coincident with SGRBs~\citep[e.g.,][]{2015ApJ...809...53C,2016JCAP...11..056P}.
High-energy neutrino emission from late prompt emission such as the extended emission has also been suggested~\citep{2006PhRvL..97e1101M}. 

Recently, a tentative $\sim3\sigma$ detection of extended \gray~emission from FRB 131104 was reported~\citep{DeLaunay+16frb131104}.  An interesting indication is that the observed \gray~emission is similar to the extended emission observed in SGRBs. Needless to say, future confirmations of such phenomena are necessary to conclude that a fraction of the FRBs is indeed accompanied by \gray s. Our work here has demonstrated that the constraints that can be placed by such \gray~observations are indeed powerful, which may require serious revisions of the extragalactic models, or the revisiting of Galactic models in which energetics and compactness issues are much less serious.  We also have shown the implications of late-time radio observations by \cite{sr16}, but they are not strong enough to test the merger scenario.  Any future firm detection of \gray s with {\it Swift}, {\it Fermi}, and other all-sky \gray~monitors, as well as late follow-up observations in the radio band, could confirm the inference presented here that FRBs originate from catastrophic cosmological events, and future searches with multi-messenger networks such as the Astrophysical Multimessenger Observatory Network (AMON)~\citep{2013APh....45...56S} could also be very important.

\begin{acknowledgements}
We thank Maurcio Bustamante, Ke Fang, Kazumi Kashiyama, and Shuta Tanaka for useful discussions.
We acknowledge support from NSF Grant No.~PHY-1620777 (KM) and NASA Grant No.~NNX13AH50G (PM).
\end{acknowledgements}

\bibliographystyle{apj_8}
\bibliography{ref}

\end{document}